\documentclass{article}
\usepackage{amsmath,amssymb,pxfonts}
\title{Short-time at-the-money skew and rough fractional volatility\thanks{This work was supported by Japan Society for the Promotion of Science under KAKENHI Grant Number 24684006.}}
\author{Masaaki Fukasawa\thanks{ Department of Mathematics, Osaka University,  1-1 Machikaneyama, Toyonaka, Osaka, JAPAN
 Email: fukasawa@math.sci.osaka-u.ac.jp}
}
\newtheorem{lem}{Lemma}
\newtheorem{thm}{Theorem}
\newtheorem{rem}{Remark}
\begin{document}
\maketitle
\begin{abstract}
 The Black-Scholes implied volatility skew at the money of SPX options is known to obey a power law with respect to the time-to-maturity. We construct a model of the underlying asset price process which is dynamically consistent to the power law.
 The volatility process of the model is driven by a fractional Brownian motion with Hurst parameter less than half.
 The fractional Brownian motion is correlated with a Brownian motion which drives the asset price process.
 We derive an asymptotic expansion of the implied volatility as the time-to-maturity tends to zero. For this purpose we introduce a new approach to validate such an expansion, which enables us to treat more general models than in the literature. The local-stochastic volatility model is treated as well under an essentially minimal regularity condition in order to show such a standard model cannot be dynamically consistent to the power law.
\end{abstract}
\section{Introduction}
The Black-Scholes implied volatility is a nonlinearly transformed price of
a call or put option in such a way that the transformed value does not depend on the strike price and the maturity of the option only if the underlying asset price is log-normally distributed under the pricing measure.
As a function of strike price and maturity,
 the implied volatilities form a surface which visually characterizes the marginal distributions of the underlying asset price under the pricing measure. In particular, it gives an idea how the price dynamics deviates from the Black-Scholes model. Its overall level tells how the underlying asset is risky. 
 The implied volatility surface of SPX option prices is usually not flat and typically exhibits downward slope and convexity as a function of the log-strike price.
 Further, it has been reported that the slope around at-the-money obeys a power law  with respect to the time-to-maturity: see Al\'os et al.~\cite{Alos}, Fouque et al.~\cite{FPSS}, Gatheral et al.~\cite{GJR}.
 Denoting by  $\sigma_t(k, \theta)$  the implied volatility at time $t$ with log-moneyness $k$ and time-to-maturity $\theta$,
 the power law can be formulated as
 \begin{equation}\label{ts}
 \frac{\sigma_t(\sqrt{\theta}z, \theta) - \sigma_t(\sqrt{\theta}\zeta,\theta)}{\sqrt{\theta} (z-\zeta)}
  \sim A_t \theta^{H-1/2}
  \ \ \text{a.s.}
 \end{equation}
 as $\theta \to 0$ for $z \neq \zeta$ and $t \geq 0$, 
 where $H \in (0,1/2)$ and $A$ is a stochastic process.
The left hand side is essentially $\partial _k \sigma_t(0,\theta)$. The above finite difference form is more relevant here because only a finite number of strike prices are listed in real markets.
 The aim of this study is to construct a model which yields (\ref{ts}).
 
 Needless to say, a financial practice needs a model being consistent to the implied volatility surface.
 A popular approach for practitioners is to model the underlying asset price $S$ under the pricing measure  by a local volatility model
\begin{equation} \label{lv}
 \mathrm{d}S_t = S_t v(S_t,t)\mathrm{d}B_t,
\end{equation}
where $v$ is a Borel function and $B$ is a standard Brownian motion.
As shown by Dupire~\cite{Dupire}, for any arbitrage-free set of 
vanilla option prices,
there exists a function $v$ such that each of the given option prices coincides with
 the theoretical no-arbitrage price under (\ref{lv}) for the corresponding payoff.
 The procedure of finding such a function $v$ given a set of market prices,
 called the calibration, has been the first step to price exotic derivatives without providing a static arbitrage opportunity.

A more important practice is to hedge an option portfolio.
The local volatility model is  not satisfactory for this purpose due to
the lack of dynamic consistency;
the calibration to market prices at different times usually gives different functions as $v$.
As a result, an hedging strategy under a model calibrated at time $t$
is outdated at time $s > t$.
This simply implies that the underlying asset price $S$  in fact does not satisfy (\ref{lv}).
The hedging error is  then out of control, at least from theoretical point of view.

The necessity of the re-calibration can be deduced from the lack of dynamic consistency to the power law.
As shown in Section~2, under a local-stochastic volatility model extending (\ref{lv}):
\begin{equation*}
  \begin{split}
   & \mathrm{d}S_t = S_t v(S_t,Y_t,t) \mathrm{d}B_t,\\
   & \mathrm{d}Y^i_t = b^i(S_t,Y_t,t) \mathrm{d}t + \sum_{j=1}^k c^i_j(S_t,Y_t,t)\mathrm{d}W^j_t, \ \
   i = 1,\dots,d,
  \end{split}
\end{equation*}
where $W= (W^1,\dots,W^k)$ is a $k$-dimensional standard Brownian motion with
$\mathrm{d}\langle B, W^i \rangle_t = \rho^i(S_t,Y_t,t)\mathrm{d}t$, we have
\begin{equation}\label{reg}
 \frac{\sigma_t(\sqrt{\theta}z, \theta) - \sigma_t(\sqrt{\theta}\zeta,\theta)}{\sqrt{\theta} (z-\zeta)}
  \to \frac{1}{2} \left\{ S_t \partial_sv(S_t,Y_t,t)
		   + ((c\rho) \cdot \nabla_y \log v) (S_t,Y_t,t)\right\}
  \ \ \text{a.s.}
\end{equation}
as $\theta \to 0$ for all $t$ and $z \neq \zeta$.
 This result partially extends Medvedev and Scaillet~\cite{MS} and Osajima~\cite{Osa1}.
 The point here is that (\ref{reg}) holds under an essentially minimal regularity condition.
This implies that
the local volatility model (\ref{lv}) needs a volatility function $v(s,t)$ which is singular
at $(s,t) = (S_t,t)$ in order to be consistent to the power law at time $t$.
To be dynamically consistent, the model needs a volatility function $v(s,t)$ which is singular everywhere and this is nonsense.
 
 As an application of general theories,
 Al\'os et al.~\cite{Alos} and Fukasawa~\cite{F1} treated volatility processes driven by a fractional Brownian motion to 
find that the term structure (\ref{ts}) at fixed time $t$ follows 
under those specific stochastic volatility  models.
The Hurst parameter $H$ of the fractional Brownian motion has to
be chosen from $(0,1/2)$ to match (\ref{ts}).
Therefore the volatility is not a process of long memory.
See Gatheral et al.~\cite{GJR} for an empirical work which suggests that the volatility appears in fact a fractional Brownian motion with $H \in (0,1/2)$.
The models of Al\'os et al.~\cite{Alos} and Fukasawa~\cite{F1} are however not dynamically consistent to (\ref{ts}) in the sense that
the models have to depend on $t$ to yield (\ref{ts}).
Therefore it suffers from the same drawback as local volatility models do.
 Due to the fact that the fractional Brownian motion is not Markov, 
the construction of a dynamically consistent model is not a trivial exercise.
In Section~3,
we use a representation of a fractional Brownian motion given by Muravlev~\cite{Mur} to solve the problem
 and show that the constructed model in fact yields (\ref{ts}) for all $t$.
 More precisely, we have (\ref{ts}) for all $t$ with
 \begin{equation} \label{Ac}
  A_t = c \partial \log v(Y_t)
 \end{equation}
under
\begin{equation}\label{rfv}
\begin{split}
 & \mathrm{d}S_t = S_t v(Y_t) \mathrm{d}B_t,\\
 & Y_t = Y_0 + \int_0^tb(Y_u)\mathrm{d}u + W^H_t,
\end{split}
\end{equation}
where $W^H$ is a fractional Brownian motion with Hurst parameter $H \in (0,1/2)$ with correlation
\begin{equation} \label{cor}
 E[(B_{t +\theta}-B_t)(W^H_{t+\theta}-W^H_t)|\mathcal{F}_t] =
c^\prime \theta^{H+1/2}
\end{equation}
and $c$, $c^\prime$ are constants.
  The potential usefulness of the Muravlev representation in finance was discussed by Novikov~\cite{Nov}.

  Now we explain how our asymptotic analysis is related to other approaches in the literature.
 There are two major categories. 
 The first one is based on a perturbation from the Black-Scholes model.
 The idea is to introduce an artificial perturbation parameter in, say, a stochastic volatility model in such a way that the model converges to the Black-Scholes model in a suitable sense as the perturbation parameter tends to zero.
 The implied volatility then converges to the volatility parameter of the limit Black-Scholes model. An asymptotic expansion of the implied volatility is then derived around the limit volatility parameter.
 The small vol-of-vol expansion by Lewis~\cite{Lewis}, the singular perturbation (fast mean reverting) expansion by Fouque et al~\cite{Fou1} and
 the multi-scale expansion  by Fouque et al~\cite{Fou2}
 belong to this category and can be verified in a unified manner as in Fukasawa~\cite{F1}.
 The second category is based on the short-time behavior of the underlying asset price process.
 There are again two major approaches within this category.
 The first one considers the implied volatility in the original scale of the strike price.  The large deviation principle and the heat kernel expansion are  explicitly or implicitly underlying this approach.
 See e.g., Berestycki et al~\cite{BBF}, Osajima~\cite{Osa1, Osa2},  Henry-Labord\'ere~\cite{HL},
 Pham~\cite{Pham}, Forde and Jacquier~\cite{FJ}, Gatheral et al~\cite{GW} and Armstrong et al.~\cite{Arm}.
 The second one rescales the strike price to get a high resolution around at-the-money.
 See  Yoshida~\cite{Y}, Kunitomo and Takahashi~\cite{KT},
 Medvedev and Scaillet~\cite{MS}, Osajima~\cite{Osa1} and Mijatovi\'c and Tankov~\cite{MT}.
 This last approach is the most relevant here because we are considering the term structure of the implied volatility around at-the-money. As already mentioned, (\ref{reg}) partially extends  Medvedev and Scaillet~\cite{MS} and Osajima~\cite{Osa1}.
 The former is based on a formal expansion of the PDE that the implied volatility satisfies.
 The latter, as well as  Yoshida~\cite{Y}, Kunitomo and Takahashi~\cite{KT},
 is based on the Watanabe theory of the Malliavin calculus.
 Our new method requires less regularity conditions and is effective to derive (\ref{ts}) under (\ref{rfv}).
 We do not take the jumps of the asset price process into account because
 Mijatovi\'c and Tankov~\cite{MT} has already considered an exponential L\'evy model to find that the implied volatility behaves differently  from (\ref{ts}).
 
 We conclude this section with an additional remark on the  jumps. By a seminal work by A$\ddot{\i}$t-Sahalia and Jacod~\cite{AJ}, it has been widely believed that continuous asset price models are rejected by statistical testing. In fact,
 A$\ddot{\i}$t-Sahalia and Jacod~\cite{AJ} and other related studies have always assumed that the volatility is an It$\hat{\text{o}}$ process.
 In other words, what has been rejected is only
 a continuous asset price model with volatility being an It$\hat{\text{o}}$ process.
 The rough fractional volatility in (\ref{rfv}) is not an It$\hat{\text{o}}$ process.
 It remains for future research to develop a statistical theory for such a model based on high frequency data.
 In this direction, as already mentioned,
 Gatheral et al.~\cite{GJR} found that modeling the volatility with $W^H$, $H\in(0,1/2)$
 is consistent to a scaling law observed in high frequency volatility times series.

\section{The local stochastic volatility model}
Here we study the short-term behavior of at-the-money skew under a regular
local stochastic volatility model extending (\ref{lv}).
Let $(\Omega,\mathcal{F},P, \{\mathcal{F}_t\}_{t\geq 0})$ be an filtered probability space satisfying the usual conditions.
We suppose a Markov structure under the pricing measure:
\begin{equation*}
  \begin{split}
   & \mathrm{d}S_t = S_t v(S_t,Y_t,t) \mathrm{d}B_t,\\
   & \mathrm{d}Y^i_t = b^i(S_t,Y_t,t) \mathrm{d}t + \sum_{j=1}^k c^i_j(S_t,Y_t,t)\mathrm{d}W^j_t, \ \
   i = 1,\dots,d,
  \end{split}
\end{equation*}
where $B$ is an $\{\mathcal{F}_t\}$-standard Brownian motion and $W= (W^1,\dots,W^k)$ is a $k$-dimensional $\{\mathcal{F}_t\}$-standard Brownian motion with
$\mathrm{d}\langle B, W^i \rangle_t = \rho^i(S_t,Y_t,t)\mathrm{d}t$.
The continuous functions $v$, $b^i$, $c^i_j$ and $\rho^i$ are defined on
$(0,\infty) \times \mathbb{R}^d \times [0,\infty)$.
Denote 
\begin{equation*}
 a^{ij} = \sum_{l=1}^k c^i_l c^j_l, \ \
  \eta^i = \sum_{l=1}^k c^i_l \rho^l, \ \
  \partial_s = \frac{\partial}{\partial s}, \ \ 
 \nabla_y  = \left(
	     \frac{\partial }{\partial y^1},
	     \dots,
	     \frac{\partial }{\partial y^d}
	    \right)
\end{equation*}
and $\eta= (\eta^1,\dots, \eta^d)$, $a=[a^{ij}]_{i,j=1}^d$.
Here we work under the following regularity conditions:
\begin{enumerate}
 \item $v(s,y,t)$ is positive,  bounded in $s$ and of linear growth in $y$
 \item  $b^i(s,y,t)$ and $c^i_j(s,y,t)$ are of linear growth in $(s,y)$.
\item 
$v(s,y,t)$ is continuously differentiable in $(s,y)$ and that
there exists $k \in \mathbb{N}$ such that
\begin{equation*}
 \sup_{s>0,y \in \mathbb{R}^d,t \geq 0}
\frac{|\partial_sv(s,y,t)| + |\nabla_y v(s,y,t)|}{1+s^k} < \infty.
\end{equation*}
 \item $v(s,y,t)$ is locally
      $H$-H$\ddot{\text{o}}$lder continuous in $t$ with $H > 1/2$.
\end{enumerate}

\begin{thm}\label{thm1}
For any $z \in \mathbb{R}$ and $t \geq 0$,
\begin{equation*}
\frac{E[(S_te^{\sqrt{\theta}z} - S_{t + \theta})_+|\mathcal{F}_t]}{S_t \sqrt{\theta}}
=
\Delta\Phi\left(\frac{\Delta}{v_t}\right)
+ (v_t +  \alpha_t \sqrt{\theta}) \phi\left(\frac{\Delta}{v_t}\right)
 + o(\sqrt{\theta}) \ \ \text{a.s.}
\end{equation*}
as $\theta \to 0$,
where
\begin{equation*}
\Delta = \frac{e^{\sqrt{\theta}z} -1}{\sqrt{\theta}}, \ \ 
v_t = v(S_t,Y_t,t)
\end{equation*}
and
\begin{equation*}
\alpha_t = \frac{z}{2}(v(S_t,Y_t,t) + S_t \partial_s v(S_t,Y_t,t) 
+ (\eta \cdot \nabla_y \log v)(S_t,Y_t,t)).
\end{equation*}
\end{thm}
{\it Proof: }
Since $(S,Y)$ is a (time-inhomogeneous) Markov process, we can and do assume $t = 0$ and 
$\mathcal{F}_0$ is trivial
without loss of generality.
Define the rescaled processes $X^\theta$ and $Y^\theta$ by  $X^\theta_u = \theta^{-1/2} (S_{\theta u} - S_0)/S_0$
and $Y^\theta_u = \theta^{-1/2}(Y_{\theta u}-Y_0)$.
The rest consists of two steps.\\

\noindent
Step 1) Here we show that
$(X^\theta_u, Y^\theta_u)$ is uniformly integrable in $\theta$ 
and  converges in law to, say, $(X^0_u,Y^0_u)$ which is
 normally distributed with
 \begin{equation*}
  \begin{split}
  &E[X^0_u] = 0, \\
   &E[Y^{0,i}_u] = 0, \\
   & E[|X^0_u|^2] = v(S_0,Y_0,0)^2 u,\\
   & E[X^0_u Y^{0,i}_u] = v(S_0,Y_0,0)\eta^i(S_0,Y_0,0)u, \\
   & E[Y^{0,i}_u Y^{0,j}_u] = a^{ij}(S_0,Y_0,0)u
 \end{split}\end{equation*}
 as $\theta \to 0$ for each $u \geq 0$.
Under the regularity conditions, an application of Gronwall's lemma gives that
\begin{equation*}
 E[\sup_{0\leq u \leq t} |S_u|^2] +
  E[\sup_{0\leq u \leq t} |Y_u|^2] <\infty
\end{equation*}
for each $t \geq 0$.
 It follows then that
\begin{equation*}
 \lim_{\theta \to 0} E[S_\theta^2 v(S_\theta,Y_\theta,\theta)^2] =
  S_0^2 v(S_0,Y_0,0)^2
\end{equation*}
and so,
\begin{equation*}
 E[|X^\theta_u|^2] = S_0^{-2}\int_0^u
  E[S_{\theta r}^2 v(S_{\theta r},Y_{\theta r},\theta r)^2]\mathrm{d}r \to
  v(S_0,Y_0,0)^2u
\end{equation*}
as $\theta \to 0$ for each $u \geq 0$.
In particular, we have that
$X^\theta_u$ is uniformly integrable in $\theta$ for each $u$.
Let $M^\theta$ be the local martingale part of $Y^\theta$.
 Since
\begin{equation*}
 E[|M^\theta_u|^2] = \sum_{i=1}^d
  \int_0^u E[|a^{ii}(S_{\theta u},Y_{\theta u},\theta u)|^2]\mathrm{d}u
  \to \sum_{i=1}^d |a^{ii}(S_0,Y_0,u)|^2 u
\end{equation*}
as $\theta \to 0$,
the vector $M^\theta_u$ is uniformly integrable
 in $\theta$ for each $u$.
Further,  the absolutely continuous part of $Y^\theta$ converges to $0$ in $L^2$.
Note also that
 \begin{equation*}
  \begin{split}
&   \langle X^\theta \rangle_u = 
   S_0^{-2}\int_0^u
  S_{\theta r}^2 v(S_{\theta r},Y_{\theta r},\theta r)^2\mathrm{d}r \to
   v(S_0,Y_0,0)^2u, \\
&     \langle X^\theta, Y^\theta \rangle_u = 
   S_0^{-1}\int_0^u
  S_{\theta r} v(S_{\theta r},Y_{\theta r},\theta r) \eta(S_{\theta r},Y_{\theta r},\theta r)\mathrm{d}r \to
   v(S_0,Y_0,0)\eta(S_0,Y_0,0) u, \\
&     \langle Y^\theta, Y^\theta \rangle_u = 
  \int_0^u
  a(S_{\theta r},Y_{\theta r},\theta r)\mathrm{d}r \to
   a(S_0,Y_0,0)u
  \end{split}
 \end{equation*}
 as $\theta \to 0$.
 Therefore the convergence in law follows from the martingale central limit theorem. \\

 \noindent
 Step 2) Define $p$ by
\begin{equation*}
 p(x,u) = (\Delta -x)
\Phi\left(
\frac{\Delta - x}{v_0\sqrt{1-u}}
\right) + v_0\sqrt{1-u}
\phi\left(
\frac{\Delta - x}{v_0\sqrt{1-u}}
\right).
\end{equation*}
This solves the partial differential equation
\begin{equation*}
 \begin{split}
&  \partial_up + \frac{1}{2}v_0^2 \partial_x^2 p = 0, \\
&  p(x,1) = (\Delta - x)_+.
 \end{split}
\end{equation*}
Note also that
\begin{equation*}
\begin{split}
& \partial_xp(x,u) = -\Phi\left(\frac{\Delta -x}{v_0\sqrt{1-u}}\right), \\ 
& \partial_x^2p(x,u) = \frac{1}{v_0\sqrt{1-u}}\phi\left(\frac{\Delta -x}{v_0\sqrt{1-u}}\right).
\end{split}
\end{equation*}
Let $f(x,y,t) = (1 + x)^2v(S_0(1+x),Y_0 + y,t)^2$.
Then, by It$\hat{\text{o}}$'s formula,
\begin{equation*}
 \begin{split}
 \frac{E[(S_0e^{\sqrt{\theta}z} - S_{\theta})_+]}{S_0 \sqrt{\theta}}
= & 
E[(\Delta - X^\theta_1)_+] \\
=& p(0,0) + \frac{1}{2}\int_0^1
E[\partial_x^2 p(X^\theta_u,u)(f(\sqrt{\theta}X^\theta_u, \sqrt{\theta}Y^\theta_u,\theta u) - v_0^2)]\mathrm{d}u.
 \end{split}
\end{equation*}
Since
\begin{equation*}
 \partial_x^2p(x,u) \to \frac{1}{v_0\sqrt{1-u}}
\phi\left(
\frac{z-x}{v_0\sqrt{1-u}}
\right)
\end{equation*}
as $\theta \to 0$, applying the result from Step 1,  we have
\begin{equation*}
 \begin{split}
&  \theta^{-1/2}
E[\partial_x^2 p(X^\theta_u,u)(f(\sqrt{\theta}X^\theta_u, \sqrt{\theta}Y^\theta_u,\theta u) - f(0, \sqrt{\theta}Y^\theta_u,\theta u))] \\
& = 
E[\partial_x^2p(X^\theta_u,u)X^\theta_u 
\int_0^1 \partial_xf(\lambda \sqrt{\theta}X^\theta_u,\sqrt{\theta}Y^\theta_u,\theta u)\mathrm{d}\lambda] \\
& \to \partial_xf(0,0,0)\frac{1}{v_0\sqrt{1-u}}
E\left[
\phi\left(\frac{z-X^0_u}{v_0\sqrt{1-u}}\right) X^0_u
\right]
=  \partial_xf(0,0,0)
\frac{zu}{v_0}\phi\left(\frac{z}{v_0}\right).
 \end{split}
\end{equation*}
Since
\begin{equation}\label{dom}
 \int_0^1 \frac{1}{\sqrt{1-u}}\mathrm{d}u < \infty,
\end{equation}
the dominated convergence theorem gives that
\begin{equation*}
\begin{split}
& \frac{1}{2}\int_0^1
E[\partial_x^2 p(X^\theta_u,u)(f(\sqrt{\theta}X^\theta_u, \sqrt{\theta}Y^\theta_u,\theta u) -
f(0, \sqrt{\theta}Y^\theta_u,\theta u) 
)]\mathrm{d}u
\\
& = \frac{z}{2}(v_0 + S_0\partial_sv(S_0,Y_0,0) )
\sqrt{\theta} \phi\left(
\frac{\Delta}{v_0}
\right) + o(\sqrt{\theta}).
\end{split}
\end{equation*}
Similarly,
\begin{equation*}
 \begin{split}
&  \theta^{-1/2}
E[\partial_x^2 p(X^\theta_u,u)(f(0, \sqrt{\theta}Y^\theta_u,\theta u) - f(0, 0, \theta u))] \\
& = 
E[\partial_x^2p(X^\theta_u,u)
\int_0^1Y^\theta_u \cdot \nabla_yf(0,\lambda\sqrt{\theta}Y^\theta_u,\theta u)\mathrm{d}\lambda] \\
& \to \frac{1}{v_0\sqrt{1-u}}
E\left[
\phi\left(\frac{z-X^0_u}{v_0\sqrt{1-u}}\right) Y^0_u \cdot \nabla_yf(0,0,0)
\right]
\\ & =  
\frac{zu}{v_0^2}\phi\left(\frac{z}{v_0}\right)
\eta(S_0,Y_0,0) \cdot \nabla_yf(0,0,0),
 \end{split}
\end{equation*}
and so,
\begin{equation*}
\begin{split}
& \frac{1}{2}\int_0^1
E[\partial_x^2 p(X^\theta_u,u)(f(0, \sqrt{\theta}Y^\theta_u,\theta u) -
f(0, 0,\theta u) 
)]\mathrm{d}u
\\
& = \frac{z}{2v_0} (\eta \cdot \nabla_yv)(S_0,Y_0,0) 
\sqrt{\theta} \phi\left(
\frac{\Delta}{v_0}
\right) + o(\sqrt{\theta}).
\end{split}
\end{equation*}
Finally, since $f(0,0,\cdot)$ is locally $H$-H$\ddot{\text{o}}$lder continuous with $H > 1/2$, 
\begin{equation*}
\frac{1}{2}\int_0^1
E[\partial_x^2 p(X^\theta_u,u)(f(0, 0,\theta u) -
f(0, 0, 0) 
)]\mathrm{d}u =
 o(\sqrt{\theta}),
\end{equation*}
which completes the proof.\hfill////\\

\noindent
The Black-Scholes price at time $t$ of a put option  with
time-to-maturity $\theta$  and 
strike price $K = S_te^{k}$
is by definition
\begin{equation*}
 P_t(k,\theta,\sigma)
= S_te^k\Phi(-d_2) - S_t\Phi(-d_1),
\end{equation*}
where $\Phi$ is the standard normal distribution function,
\begin{equation*}
 d_1 = \frac{ - k + \sigma^2\theta/2}{\sigma\sqrt{\theta}}, \ \ 
d_2 = d_1 - \sigma \sqrt{\theta}.
\end{equation*}
This is an increasing function of the parameter  $\sigma$, which is called the volatility. The Black-Scholes implied volatility
$\sigma_t(k,\theta)$ at time $t$ is defined through the equation
\begin{equation}\label{iv}
 P_t(k,\theta, \sigma_t(k,\theta)) = P_t(k,\theta),
\end{equation}
where $P_t(k,\theta)$ is the corresponding price of the put option.

\begin{thm}\label{thm2}
For any $z \in \mathbb{R}$ and $t \geq 0$,
\begin{equation*}
\begin{split}
 \sigma_t(\sqrt{\theta}z ,\theta)
= & \  v(S_t,Y_t,t)
\\ &+
(S_t\partial_sv(S_t,Y_t,y) + (\eta \cdot \nabla_y \log v)(S_t,Y_t,t))
 \frac{\sqrt{\theta}z}{2} + o(\sqrt{\theta})
  \ \ \text{a.s.}
\end{split}
\end{equation*} 
as $\theta \to 0$.
\end{thm}
{\it Proof: }
In light of Theorem~\ref{thm1}, it suffices to show that
\begin{equation*}
\frac{P_t(\sqrt{\theta}z,\theta)}{S_t \sqrt{\theta}} =
\Delta\Phi\left(\frac{\Delta}{v_t}\right)
+ (v_t +  \alpha_t \sqrt{\theta}) \phi\left(\frac{\Delta}{v_t}\right)
 + o(\sqrt{\theta})
\end{equation*}
implies
\begin{equation*}
 \sigma_t(\sqrt{\theta}z,\theta) = 
\left(1 - \frac{\sqrt{\theta}z}{2}\right)v_t + \alpha_t \sqrt{\theta} +
o(\sqrt{\theta}).
\end{equation*}
Using that $\Delta = z + \sqrt{\theta}z^2/2 + o(\sqrt{\theta})$ and
$\phi^\prime(x) = -x\phi(x)$, we have
\begin{equation*}
\frac{P_t(\sqrt{\theta}z,\theta)}{S_t \sqrt{\theta}} =
\Delta\Phi\left(\frac{z}{v_t}\right)
+ (v_t +  \alpha_t \sqrt{\theta}) \phi\left(\frac{z}{v_t}\right)
 + o(\sqrt{\theta})
\end{equation*}
On the other hand, by definition,
\begin{equation*}
\begin{split}
\frac{ P_t(\sqrt{\theta}z,\theta,\sigma)}{S_t \sqrt{\theta}}
& \ = \Delta \Phi(-d_2) + \frac{\Phi(-d_2)-\Phi(-d_1)}{\sqrt{\theta}} \\
& \ = 
\Delta 
\Phi\left(\frac{z}{\sigma}\right) + \phi
\left(\frac{z}{\sigma}\right) \frac{\sigma}{2}\sqrt{\theta} z +
\sigma \phi
\left(\frac{z}{\sigma}\right) + o(\sqrt{\theta}).
\end{split}
\end{equation*}
It is then easy to see $\sigma_t(\sqrt{\theta}z,\theta) = v_t + o(1)$.
Put $\beta(\theta) = \sigma_t(\sqrt{\theta}z,\theta)-v_t$. Then,
\begin{equation*}
\begin{split}
&\frac{ P_t(\sqrt{\theta}z,\theta,v_t + \beta(\theta))}{S_t \sqrt{\theta}}
  \\ & = 
\Delta 
\Phi\left(\frac{z}{v_t}\right) + \phi
\left(\frac{z}{v_t}\right) \frac{v_t}{2}\sqrt{\theta} z +
(v_t + \beta(\theta)) \phi
\left(\frac{z}{v_t}\right) + o(\sqrt{\theta}) +
o(\beta(\theta))
\end{split}
\end{equation*}
and so,
\begin{equation*}
 \alpha_t \sqrt{\theta} = \frac{v_t}{2} \sqrt{\theta}z + \beta(\theta)
 + o(\sqrt{\theta}) +
o(\beta(\theta)).
\end{equation*}
Therefore,
\begin{equation*}
 \theta^{-1/2}\beta(\theta) \to \alpha_t - \frac{v_t}{2}z, 
\end{equation*}
which completes the proof. \hfill////\\

\noindent
We have seen that if $z \neq \zeta$,
\begin{equation*}
 \frac{\sigma_t(\sqrt{\theta}z, \theta) - \sigma_t(\zeta,\theta)}{\sqrt{\theta} (z-\zeta)}
  \to \frac{1}{2} \left\{ S_t \partial_sv(S_t,Y_t,t)
		   + (\eta \cdot \nabla_y \log v) (S_t,Y_t,t)\right\}
  \ \ \text{a.s.}
\end{equation*}
as $\theta \to 0$ under a regular local-stochastic volatility model.
This means that the regular models cannot explain
the empirically observed term structure of the implied volatility skew (\ref{ts}).
As a result, a forcible calibration at time $t$
leads to a function $v$ which has a kind of singularity at
$(s,y,t) = (S_t,Y_t,t)$.
Therefore, it has to be singular everywhere in order to have (\ref{ts}) for all $t$.
It is hopeless and even nonsense to have such a singular function by any practical method of calibration.

\section{Rough stochastic volatility model}
Here, we consider a model with stochastic volatility being rough. More precisely, we model the volatility to be driven by a fractional Brownian motion with Hurst parameter $H \in (0,1/2)$.
We show that the term structure (\ref{ts}) follows when the fractional Brownian motion is correlated with the Brownian motion that drives the underlying asset price process.
Let $(\Omega,\mathcal{F},P, \{\mathcal{F}_t\}_{t\geq 0})$ be an filtered probability space satisfying the usual conditions.
The underlying asset price model under the pricing measure is
\begin{equation*}
\begin{split}
 & \mathrm{d}S_t = S_t v(S_t,Y_t,t) \mathrm{d}B_t,\\
 & Y_t = Y_0 + \int_0^tb(Y_u)\mathrm{d}u + W^H_t,
\end{split}
\end{equation*}
where $B$ is an $\{\mathcal{F}_t\}$-standard Brownian motion, 
$W^H$ is a fractional Brownian motion with Hurst parameter $H \in (0,1/2)$.
Here we work under the following regularity conditions:
\begin{enumerate}
 \item $v(s,y,t)$ is positive, bounded in $s$ and of linear growth in $y$.
\item 
$v(s,y,t)$ is continuously differentiable in $(s,y)$ and that
there exists $k \in \mathbb{N}$ such that
\begin{equation*}
 \sup_{s>0,y \in \mathbb{R},t \geq 0}
\frac{|\partial_sv(s,y,t)| + |\partial_y v(s,y,t)|}{1+s^k} < \infty.
\end{equation*}
 \item for each $(s,y)$, there exists $\epsilon > 0$ such that
      $v(s,y,t)$ is locally 
       $(H+\epsilon)$-H$\ddot{\text{o}}$lder continuous in $t$.
 \item for each $s$,
       there exists $\epsilon > 0$ such that
      $\partial_y v(s,y,t)$ is locally
       $\epsilon$-H$\ddot{\text{o}}$lder continuous in $(y,t)$.
       \item  $b$ is Lipschitz continuous.
\end{enumerate}
To introduce the correlation between $B$ and $W^H$, we adopt a representation of the fractional Brownian motion given by Muravlev~\cite{Mur}:
 \begin{equation*}
\begin{split}
& W^H_t = c_H \int_0^\infty \beta^{-1/2-H}(Z^\beta_t-Z^\beta_0) \mathrm{d}\beta, \\
& Z^\beta_t = \int_{-\infty}^t e^{-\beta(t-s)}\mathrm{d}W_s,
\end{split}
\end{equation*}
where $c_H>0$ is a constant and $W$ is an $\{\mathcal{F}_t\}$-standard Brownian motion.
We assume $\mathrm{d}\langle B,W \rangle_t = \rho(Y_t)\mathrm{d}t$
with a continuous function $\rho$ with $|\rho|\leq 1$.

\begin{thm}\label{thm3}
For any $z \in \mathbb{R}$ and $t \geq 0$,
\begin{equation*}
\frac{E[(S_te^{\sqrt{\theta}z} - S_{t + \theta})_+|\mathcal{F}_t]}{S_t \sqrt{\theta}}
=
\Delta\Phi\left(\frac{\Delta}{v_t}\right)
+ (v_t +  \alpha^\theta_t \theta^H) \phi\left(\frac{\Delta}{v_t}\right)
 + o(\theta^H)     \ \ \text{a.s.}
\end{equation*}
as $\theta \to 0$,
where
\begin{equation*}
\Delta = \frac{e^{\sqrt{\theta}z} -1}{\sqrt{\theta}}, \ \ 
 v_t = v(S_t,Y_t,t)
\end{equation*}
 and
 \begin{equation} \label{alpt}
  \begin{split}
 &\alpha^\theta_t =
   \frac{c_H\Gamma(1/2-H)}{(1/2+H)(3/2+H)} z \rho(Y_t)\partial_y \log v(S_t,Y_t,t) + F^\theta_t \partial_y  v(S_t,Y_t,t),\\
   &
   F^\theta_t = \int_0^\infty \beta^{-3/2-H}(1-\beta-e^{-\beta}) \theta^{-1/2}Z^{\beta/\theta}_t \mathrm{d}\beta.
  \end{split}
  \end{equation}
\end{thm}
\begin{rem}
 $F^\theta_t$ is a functional of $\{W_s\}_{-\infty < s \leq t}$.
 The law of $F^\theta_t$ does not depend on $\theta$.
\end{rem}
From Theorem~\ref{thm3}, by the same argument as in
 the proof of Theorem~\ref{thm2},
 we obtain the main result of this paper:

 \begin{thm}
  For any $z \in \mathbb{R}$ and $t \geq 0$,
\begin{equation*}
 \sigma_t(\sqrt{\theta}z ,\theta)
=   v(S_t,Y_t,t) +  \alpha^\theta_t \theta^H + o(\theta^H)
  \ \ \text{a.s.}
\end{equation*} 
  as $\theta \to 0$, where $\alpha^\theta_t$ is defined by (\ref{alpt}).
  In particular, if $z \neq \zeta$,
  \begin{equation*}
\frac{   \sigma_t(\sqrt{\theta}z ,\theta)
 - \sigma_t(\sqrt{\theta}\zeta ,\theta)}
 {\sqrt{\theta}(z-\zeta)}
 \sim 
 \frac{c_H\Gamma(1/2-H)}{(1/2+H)(3/2+H)}  \rho(Y_t)\partial_y \log v(S_t,Y_t,t)\theta^{H-1/2}
 \ \ \text{a.s.}
  \end{equation*}
  as $\theta \to 0$.
When $\rho$ is constant and $v(s,y,t)=v(y)$, we have (\ref{ts}) with (\ref{Ac}) and (\ref{cor}).
 \end{thm}
 We start with some lemmas.
\begin{lem} \label{lem1}
  For every $t, s \in \mathbb{R}$,
 \begin{equation*}
  \int_0^\infty \beta^{-1/2 - H}|Z^\beta_t - Z^\beta_s|\mathrm{d}\beta < \infty, \ \ 
    \int_0^\infty \beta^{-1/2 - H}|e^{-\beta(t-s)}-1||Z^\beta_s|\mathrm{d}\beta < \infty \ \ \text{a.s.}
 \end{equation*}
  and
  \begin{equation*}
    \begin{split}
   W^H_t  = & \ W^H_s +
     c_H      \int_0^\infty \beta^{-1/2 - H}(Z^\beta_t - Z^\beta_s)\mathrm{d}\beta \\
      = & \ W^H_s + c_H
     \int_0^\infty \beta^{-1/2 - H} (e^{-\beta(t-s)}-1)  Z^\beta_s\mathrm{d}\beta \\ &+
     c_H  \int_0^\infty \beta^{-1/2 - H} \int_s^t e^{-\beta(t-u)}\mathrm{d}W_u\mathrm{d}\beta
     \ \ \text{a.s..}
     \end{split}
  \end{equation*}
 \end{lem}
 {\it Proof: }
Let $t > s$ without loss of generality. By definition,
  \begin{equation*}
   Z^\beta_t = e^{\beta(s-t)}Z^\beta_s + \int_s^t e^{-\beta(t-u)}\mathrm{d}W_u
  \end{equation*}
  and so,
  \begin{equation*}
   \begin{split}
   E[|Z^\beta_t - Z^\beta_s|^2]
    & \leq   2 |e^{\beta(s-t)}-1|^2 E[|Z^\beta_s|^2] + 2
    \int_s^t e^{-2\beta(t-u)}\mathrm{d}u \\
    & = \beta^{-1}|e^{\beta(s-t)}-1|^2 + \beta^{-1}(1-e^{2\beta(s-t)}) \\
    & = 2 \beta^{-1}(1 - e^{-\beta(t-s)}) \\
    & \leq 2 \min\{(t-s), \beta^{-1}\}.
   \end{split}
  \end{equation*}
  Therefore,
  \begin{equation*}
   E\left[
  \int_0^\infty \beta^{-1/2 - H}|Z^\beta_t - Z^\beta_s|\mathrm{d}\beta
	   \right]  
   \leq 
    \int_0^\infty \beta^{-1/2 - H}E[|Z^\beta_t - Z^\beta_s|^2]^{1/2}\mathrm{d}\beta < \infty.
  \end{equation*}
  The rest is obvious. \hfill////
  \begin{lem}\label{lem2}
For all $t \in \mathbb{R}$, $\theta \geq 0$ and for all Borel function $f$, 
   \begin{equation*}
E[f(S_{t + \theta},Y_{t + \theta}) | \mathcal{F}_t]
 = E[f(S_{t + \theta},Y_{t+\theta})| S_t, Y_t, \{Z^\beta_t\}_{\beta > 0}]
 \ \ \text{a.s.}
   \end{equation*}
   and
   \begin{equation*}
   \begin{split}
&E[f(S_{t + \theta},Y_{t + \theta})| S_t=s, Y_t = y,  Z^\beta_t=z^\beta, \beta > 0]
\\ &=
    E[f(S_{\theta},Y_{\theta})| S_0=s, Y_0 = y, Z^\beta_0=z^\beta, \beta > 0]
        \ \ \text{a.s..}
\end{split}
\end{equation*}
  \end{lem}
  {\it Proof: }
  Since
  \begin{equation*}
   \begin{split}
    Y_u =& Y_t + \int_t^u b(Y_r)\mathrm{d}r + W^H_u-W^H_t\\
    = & Y_t + \int_t^u b(Y_r)\mathrm{d}r +
    c_H
    \int_0^\infty \beta^{-1/2 - H} e^{-\beta(u-t)}  Z^\beta_t\mathrm{d}\beta \\
     & +
c_H  \int_0^\infty \beta^{-1/2 - H} \int_t^u e^{-\beta(u-r)}\mathrm{d}W_r\mathrm{d}\beta 
   \end{split}
  \end{equation*}
  by Lemma~\ref{lem1},
  the conditional law of  $\{Y_u\}_{u \geq t}$ given
  $S_t, Y_t, \{Z^\beta_t\}_{\beta > 0}$ is independent of $\mathcal{F}_t$.
  Therefore for any bounded $\mathcal{F}_t$-measurable random variable $A$,
  \begin{equation*}
 \begin{split}
   &E[f(S_{t + \theta},Y_{t+\theta})A | S_t, Y_t, \{Z^\beta_t\}_{\beta > 0}]
    \\ &=
    E[f(S_{t + \theta},Y_{t+\theta}) | S_t, Y_t, \{Z^\beta_t\}_{\beta > 0}]
    E[A| S_t, Y_t, \{Z^\beta_t\}_{\beta > 0}]     \ \ \text{a.s..}
 \end{split}
  \end{equation*}
  It follows then that
  \begin{equation*}
   E[f(S_{t + \theta},Y_{t+\theta})A ]
    = E[E[f(S_{t + \theta},Y_{t+\theta}) | S_t, Y_t, \{Z^\beta_t\}_{\beta > 0}]A],
  \end{equation*}
  which proves the first part.
  The second part then follows from the stationarity of the Brownian increments.
  \hfill////\\

  \noindent
  The above lemma shows that
  all the information on the history of $W_s$, $s\leq t$ is translated into the spot values $Z^\beta_t$, $\beta >0$. We may say $\beta \mapsto Z^\beta_t$ is the Laplace transform of $W_s$, $s \leq t$.
  For each $\beta> 0$, $Z^\beta$ is an OU process
\begin{equation*}
  \mathrm{d}Z^\beta_t = -\beta Z^\beta_t \mathrm{d}t + \mathrm{d}W_t
\end{equation*}
and in particular, a Markov process. Note also that the infinite dimensional process
\begin{equation*}
 (S, Y,  Z^\beta ; \beta >0)
\end{equation*}
is a time-homogeneous Markov process.\\

\noindent
{\it Proof of Theorem~\ref{thm3}: }
By Lemma~\ref{lem2}, we may and do assume without loss of generality  $t=0$
and the existence of a regular conditional probability measure $P_0$ given $\mathcal{F}_0$.
In particular, $v = v(S_0, Y_0,0)$.
Denote by $E_0$ the expectation with respect to $P_0$.
Let $X^\theta_u = \theta^{-1/2}(S_{\theta u} - S_0)/S_0$,
\begin{equation*}
 \hat{Y}^\theta_u = c_H  \theta^{-H} \int_0^{\infty} \beta^{-1/2-H} \int_0^{\theta u} e^{-\beta(\theta u - r)}\mathrm{d}W_r
  \mathrm{d}\beta
\end{equation*}
and
\begin{equation*}
 Y^\theta_u =
  \theta^{-H} \int_0^{\theta u} b(Y_r)\mathrm{d}r
  + \hat{Y}^\theta_u.
\end{equation*}
Then by Lemma~\ref{lem1}, we have the decomposition
\begin{equation*}
 \begin{split}
 Y_{\theta u } = &
  Y_0 + \int_0^{\theta u} b(Y_r)\mathrm{d}r
  + c_H \int_0^\infty \beta^{-1/2 -H} (e^{-\beta u}-1)Z^\beta_0 \mathrm{d}\beta
  + \theta^H \hat{Y}^\theta_u\\
  &  =  Y_0 
  + c_H \int_0^\infty \beta^{-1/2 -H} (e^{-\beta u}-1)Z^\beta_0 \mathrm{d}\beta
  + \theta^H Y^\theta_u.
  \end{split}
\end{equation*}
The rest of the proof consists of three steps.\\

\noindent
Step 1) Here we show that
$(X^\theta_u, Y^\theta_u)$ is uniformly integrable in $\theta$  under $P_0$ 
and  converges in law to, say, $(X^0_u,Y^0_u)$ which is
 normally distributed with
\begin{equation*}
 \begin{split}
  &E[X^0_u] = E[Y^0_u] = 0, \\
  & E[|X^0_u|^2] = v_0^2u, \\
  & E[X^0_u Y^0_u] = v_0 \rho(Y_0)  c_H\frac{\Gamma(1/2-H)}{1/2+H} u^{H+1/2}.
  \end{split}
\end{equation*}
 as $\theta \to 0$ for each $u \geq 0$.
Under the regularity conditions, an application of Gronwall's lemma gives that
\begin{equation*}
 E_0[\sup_{0\leq u \leq t} |S_u|^2] +
  E_0[\sup_{0\leq u \leq t} |Y_u|^2] <\infty
\end{equation*}
for each $t \geq 0$.
It follows then that
\begin{equation*}
 \theta^{-H} \int_0^{\theta u} b(Y_r)\mathrm{d}r \to 0
\end{equation*}
as $\theta \to 0$ in $L^2(P_0)$.
 By the scaling property of $W$,
 the law of $\hat{Y}^\theta_0$ under $P_0$ does not depend on $\theta$.
 It follows then that $(X^\theta_u, Y^\theta_u)$ is uniformly integrable under $P_0$.
Let $\hat{W}^\theta_u = \theta^{-1/2}W_{\theta u} $. Then,
we have
\begin{equation*}
 \begin{split}
  E[ \hat{W}^\theta_u \hat{Y}^\theta_u|\mathcal{F}_0] =& c_H \int_0^\infty
   \beta^{-1/2-H}\int_0^u e^{-\beta(u-r)}\mathrm{d}r \mathrm{d}\beta
  = c_H \Gamma(1/2-H) \int_0^u (u-r)^{H-1/2}\mathrm{d}r \\
  = & c_H\frac{\Gamma(1/2-H)}{1/2+H} u^{H+1/2}.
 \end{split}
\end{equation*}
Note that $B$ admits a representation
\begin{equation*}
 B_u = \int_0^u \rho(Y_r)\mathrm{d}W_r +
  \int_0^u \sqrt{1-\rho(Y_r)^2}\mathrm{d}W^\perp_r
\end{equation*}
with $(W, W^\perp)$ being a $2$-dimensional $\{\mathcal{F}_t\}$-standard Brownian motion.
Then, 
\begin{equation*}
X^\theta_u - v\rho(Y_0)\hat{W}^\theta_u - v\sqrt{1-\rho(Y_0)^2} \theta^{-1/2}W^\perp_{\theta u}\to 0
\end{equation*}
in $L^2(P_0)$ as $\theta \to 0$.
It follows that the law of
$(X^\theta_u ,Y^\theta_u)$ under $P_0$
converges weakly to the law of $(X^0_u,Y^0_u)$. \\

\noindent
Step 2)
Let $g(x,y,z,t) = (1+x)^2v(S_0(1+x),Y_0 + y+z,t)^2$
and define the function $p$ as in the proof of Theorem~\ref{thm1}.
Then, by It$\hat{\text{o}}$'s formula,
\begin{equation*}
 \begin{split}
 &\frac{E_0[(S_0e^{\sqrt{\theta}z} - S_{\theta})_+]}{S_0 \sqrt{\theta}} \\
&=  
E_0[(\Delta - X^\theta_1)_+] \\
&= p(0,0) + \frac{1}{2}\int_0^1
E_0[\partial_x^2 p(X^\theta_u,u)(g(\sqrt{\theta}X^\theta_u, \theta^HY^\theta_u, \hat{h}(\theta u), \theta u) - v_0^2)]\mathrm{d}u,
 \end{split}
\end{equation*}
where
\begin{equation*}
 \hat{h}(u) = c_H \int_0^\infty \beta^{-1/2 -H} (e^{-\beta u}-1)Z^\beta_0 \mathrm{d}\beta.
\end{equation*}
Since
\begin{equation*}
 \partial_x^2p(x,u) \to \frac{1}{v_0\sqrt{1-u}}
\phi\left(
\frac{z-x}{v_0\sqrt{1-u}}
\right)
\end{equation*}
as $\theta \to 0$, applying the result from Step 1,  we have
\begin{equation*}
 \begin{split}
&  \theta^{-1/2}
  E_0[\partial_x^2 p(X^\theta_u,u)(g(\sqrt{\theta}X^\theta_u, \theta^HY^\theta_u, \hat{h}(\theta u),
  \theta u) - g(0, \theta^HY^\theta_u,\hat{h}(\theta u),\theta u))] \\
& = 
E_0[\partial_x^2p(X^\theta_u,u)X^\theta_u 
\int_0^1 \partial_xg(\lambda \sqrt{\theta}X^\theta_u,\theta^HY^\theta_u,\hat{h}(\theta u), \theta u)\mathrm{d}\lambda] \\
& \to \partial_xg(0,0,0,0)\frac{1}{v_0\sqrt{1-u}}
E\left[
\phi\left(\frac{z-X^0_u}{v_0\sqrt{1-u}}\right) X^0_u
\right]
=  \partial_xg(0,0,0,0)
\frac{zu}{v_0}\phi\left(\frac{z}{v_0}\right).
 \end{split}
\end{equation*}
Therefore, by (\ref{dom}) and the dominated convergence theorem,
\begin{equation*}
 \begin{split}
&\frac{1}{2}\int_0^1
E_0[\partial_x^2 p(X^\theta_u,u)(g(\sqrt{\theta}X^\theta_u, \theta^HY^\theta_u,\hat{h}(\theta u),\theta u) -
g(0, \theta^HY^\theta_u,\hat{h}(\theta u),\theta u) 
  )]\mathrm{d}u \\ &
  = o(\theta^H).
 \end{split} 
\end{equation*}
Similarly,
\begin{equation*}
 \begin{split}
&  \theta^{-H}
E_0[\partial_x^2 p(X^\theta_u,u)(g(0, \theta^HY^\theta_u,\hat{h}(\theta u), \theta u) - g(0, 0,\hat{h}(\theta u), \theta u))] \\
& = 
E_0[\partial_x^2p(X^\theta_0,u)
\int_0^1Y^\theta_u \partial_yg(0,\lambda\theta^HY^\theta_u, \hat{h}(\theta u), \theta u)\mathrm{d}\lambda ] \\
& \to \frac{1}{v_0\sqrt{1-u}}
E\left[
\phi\left(\frac{z-X^0_u}{v_0\sqrt{1-u}}\right) Y^0_u 
\right]\partial_yg(0,0,0,0)
\\ &=  
c_H\frac{\Gamma(1/2-H)}{1/2+H}\frac{zu^{H+1/2}}{v_0^2}\phi\left(\frac{z}{v_0}\right)
\rho(S_0,Y_0,0) \partial_yg(0,0,0,0),
 \end{split}
\end{equation*}
and so,
\begin{equation*}
\begin{split}
& \frac{1}{2}\int_0^1
E_0[\partial_x^2 p(X^\theta_u,u)(g(0, \sqrt{\theta}Y^\theta_u, \hat{h}(\theta u),\theta u) -
g(0, 0,\hat{h}(\theta u),\theta u) 
)]\mathrm{d}u
\\
 & = c_H\frac{\Gamma(1/2-H)}{(1/2+H)(3/2+H)} z \rho(Y_0)\partial_y \log v(S_0,Y_0,0) 
\theta^H \phi\left(
\frac{\Delta}{v_0}
\right) + o(\theta^H).
\end{split}
\end{equation*}
Next, observe that
\begin{equation*} 
\begin{split}
& \frac{1}{2}\int_0^1
E_0[\partial_x^2 p(X^\theta_u,u)(g(0, 0, \hat{h}(\theta u),\theta u) -
g(0, 0, 0,\theta u) 
)]\mathrm{d}u 
 \\
 & = \frac{1}{2}\int_0^1
 E_0[\partial_x^2 p(X^\theta_u,u)]
 \int_0^1 \partial_zg(0, 0, \lambda \hat{h}(\theta u),\theta u) \mathrm{d}\lambda
  \hat{h}(\theta u)\mathrm{d}u 
 \\
 & = \frac{1}{2}\int_0^1
 \left(E_0[\partial_x^2 p(X^\theta_u,u)] -\frac{1}{v_0}\phi\left(
\frac{\Delta}{v_0}
 \right)\right) 
 \int_0^1 \partial_zg(0, 0, \lambda \hat{h}(\theta u),\theta u) \mathrm{d}\lambda
 \hat{h}(\theta u)\mathrm{d}u
 \\
 & \hspace*{1cm}+
 \frac{1}{2}\int_0^1
 \frac{1}{v_0}\phi\left(
\frac{\Delta}{v_0}
 \right)
 \int_0^1 (\partial_zg(0, 0, \lambda \hat{h}(\theta u),\theta u)-\partial_zg(0,0,0,0)) \mathrm{d}\lambda
 \hat{h}(\theta u)\mathrm{d}u
 \\
 &\hspace*{1cm}+
 \theta^H F^\theta_0 \partial_y v(S_0,Y_0,0) \phi\left(\frac{\Delta}{v_0}\right).
\end{split}
\end{equation*}
We show in the next step that
\begin{equation} \label{final1}
\int_0^1
 \left(E_0[\partial_x^2 p(X^\theta_u,u)] -\frac{1}{v_0}\phi\left(
\frac{\Delta}{v_0}
 \right)\right) 
 \int_0^1 \partial_zg(0, 0, \lambda \hat{h}(\theta u),\theta u) \mathrm{d}\lambda
 \hat{h}(\theta u)\mathrm{d}u =o(\theta^H)
\end{equation}
and
\begin{equation} \label{final2}
\int_0^1
 \frac{1}{v_0}\phi\left(
\frac{\Delta}{v_0}
 \right)
 \int_0^1 (\partial_zg(0, 0, \lambda \hat{h}(\theta u),\theta u)-\partial_zg(0,0,0,0)) \mathrm{d}\lambda
 \hat{h}(\theta u)\mathrm{d}u = o(\theta^H)
\end{equation}
in the almost sure sense. Then, the proof is completed by noting that
\begin{equation*}
\frac{1}{2}\int_0^1
E_0[\partial_x^2 p(X^\theta_u,u)(g(0, 0, 0,\theta u) -
g(0,0, 0, 0) 
)]\mathrm{d}u =
 o(\theta^H)
\end{equation*}
since $g(0,0,0,\cdot)$ is locally $(H+\epsilon)$-H$\ddot{\text{o}}$lder continuous with $\epsilon > 0$.\\

\noindent
Step 3)
Here we show (\ref{final1}) and (\ref{final2}).
Let
\begin{equation*}
 h(u) = c_H \int_0^\infty \beta^{-1/2 -H} |e^{-\beta u}-1||Z^\beta_0| \mathrm{d}\beta.
\end{equation*}
Let us show that
\begin{equation} \label{convep}
 \theta^{-H + \epsilon} h(\theta) \to 0 \ \ \text{a.s.}
\end{equation}
as $\theta \to 0$ for any $\epsilon > 0$. 
Since
\begin{equation*}
 \sup_{0\leq u \leq 1} |\hat{h}(\theta u)| \leq h(\theta)
\end{equation*}
and $\partial_zg(0,0,\cdot,\cdot)$ is locally $\epsilon$-H$\ddot{\text{o}}$lder continuous,
(\ref{final2}) follows from (\ref{convep}).
As we have already seen in the proof of Lemma~\ref{lem1},
$E[h(u)] < \infty$. Further, by changing variable as $\gamma = \theta \beta$,
\begin{equation*}
 \theta^{-H}h(\theta )
  = c_H\int_0^\infty \gamma^{-1/2-H}|e^{-\gamma }-1|
  \left|\int_{-\infty}^0e^{\gamma v}\mathrm{d}\hat{W}_v^{\theta} \right| \mathrm{d}\gamma,
\end{equation*}
which means that the law of $\theta^{-H}h(\theta )$ does not depend on $\theta$.
Therefore,
\begin{equation*}
 E[ \theta^{-H + \epsilon} h(\theta )] = C \theta^\epsilon
\end{equation*}
with a constant $C>0$.
Let $\theta_n = n^{-2/\epsilon}$. Then, for any $\delta >0$
\begin{equation*}
 \sum_{n=1}^\infty P( |\theta_n^{-H+\epsilon}h(\theta_n )| > \delta)
  \leq C \sum_{n=1}^\infty \frac{1}{\delta n^2 } < \infty
\end{equation*}
and so, 
\begin{equation*}
  \theta_n^{-H + \epsilon} h(\theta_n ) \to 0 \ \ \text{a.s.}
\end{equation*}
by the Borel-Cantelli lemma.
This implies (\ref{convep}) because
 $h$ is a decreasing function and $\theta_n/\theta_{n+1} \to 1$.
 Now, notice that $ q(x,r) = \partial_x^2p(x,r)$  solves
 the partial differential equation
 \begin{equation*}
  \partial_r q + \frac{1}{2}v_0^2 \partial_x^2 q, \ \
   q(0,0) = \frac{1}{v_0}\phi\left(
\frac{\Delta}{v_0}\right)
 \end{equation*}
 and so, by It$\hat{\text{o}}$'s formula,
 \begin{equation*}
  \begin{split}
  &E_0[\partial_x^2 p(X^\theta_u,u)]  - \frac{1}{v_0}\phi\left(
   \frac{\Delta}{v_0} \right)
   \\ & = 
  \frac{1}{2}\int_0^u E_0[
  \partial_x^4p(X^\theta_r,r)
   (g(\sqrt{\theta}X^\theta_r, \theta^HY^\theta_r, \hat{h}(\theta r), \theta r) - v_0^2)]\mathrm{d}r.
  \end{split}
 \end{equation*}
 Repeating the same argument as Step 2, with the aid of (\ref{convep}), we have that
 \begin{equation*}
  \sup_{0 \leq u \leq 1} \left|
   E_0[\partial_x^2 p(X^\theta_u,u)]  - \frac{1}{v_0}\phi\left(
   \frac{\Delta}{v_0} \right) \right| = o(\theta^\delta)
 \end{equation*}
 for any $\delta \in (0,H)$.
 This and (\ref{convep}) imply (\ref{final1}). \hfill////

\end{document}